\documentclass[10pt,a4paper]{iopart}
\usepackage[utf8]{inputenc}
\usepackage[english]{babel}
\usepackage[T1]{fontenc}

\usepackage{amsfonts}
\usepackage{graphicx}
\usepackage{dcolumn}
\usepackage{bm}

\usepackage{amssymb}
\usepackage{bm}

\expandafter\let\csname equation*\endcsname\relax
\expandafter\let\csname endequation*\endcsname\relax
\usepackage{amsmath}
\usepackage{mathrsfs}
\usepackage[dvipsnames]{xcolor}

\usepackage[numbers,sort&compress]{natbib}
\usepackage{doi}

\usepackage{hyperref}
\hypersetup{
	colorlinks=true,
	linkcolor=blue,
	anchorcolor=blue,
    urlcolor=blue,
	citecolor=blue}

\let\originalleft\left
\let\originalright\right
\renewcommand{\left}{\mathopen{}\mathclose\bgroup\originalleft}
\renewcommand{\right}{\aftergroup\egroup\originalright}

\newcommand{\mga}[1]{{#1}}
\newcommand{\cyan}[1]{{#1}}

\newcommand{\FA}[1]{{#1}}
\begin{document}











\title[Quantifying complexity of continuous-variable quantum states]{Quantifying complexity of continuous-variable quantum states via Wehrl entropy and Fisher information}

\author{Siting Tang$^{1,2,3}$, Francesco Albarelli$^{4}$,  Yue Zhang$^{1,2}$, Shunlong Luo$^{1,2}$, and Matteo G. A. Paris$^{3}$}
\address{$^{1}$State Key Laboratory of Mathematical Sciences, Academy of Mathematics and Systems Science, Chinese Academy of Sciences,  Beijing 100190,  China}
\address{$^{2}$School of Mathematical Sciences, University of Chinese Academy of Sciences, Beijing 100049, China}
\address{$^{3}$ Dipartimento di Fisica, Universit\`a  di Milano, I-20133 Milano, Italy}
\address{$^{4}$Scuola Normale Superiore, I-56126 Pisa, Italy}


\vspace{10pt}
\begin{indented}
\item[] \today
\end{indented}

\vskip 0.5cm

\begin{abstract}
The notion of complexity of quantum states is quite different from uncertainty or information contents, and involves the tradeoff between its classical and quantum features. 
In this work, we 
\FA{we introduce a quantifier of complexity of continuous-variable states, e.g. quantum optical states, based on the Husimi quasiprobability distribution.
This quantity is built upon two functions of the state: the Wehrl entropy, capturing the spread of the distribution, and the Fisher information with respect to location parameters, which captures the opposite behaviour, i.e. localization in phase space.}
We analyze the basic properties of the quantifier and illustrate its features by evaluating complexity
of Gaussian states and some relevant non-Gaussian states. We further generalize the quantifier in terms of $s$-ordered phase-space distributions and illustrate its implications.
\end{abstract}




\section{Introduction}

Literally, the word ``complexity'' refers to the degree to which an object deviates from being ``simple'' or ``ideal''. However, its meaning can vary significantly among different scientific disciplines, such as physics~\cite{Ande1991}, biology~\cite{Pari1993}, mathematics~\cite{Baba1984}, information theory~\cite{Kolm1965}, and computer science~\cite{Watr2009,Aaro2016}.
Complexity of quantum systems is a ubiquitous notion with wide implications and applications~\cite{Sen2011,Bern1997,Kemp2006,Gonz2009,Lopez-Ruiz1995,Sanu2008,Catalan2002,Romera2004,Shiner1999,Soko2008,Bene2009,Bala2010,Bene2012,Manzano2012,Qin2014,Calb2001,Dehe2009,Brenner2024}, most of which are related to the computational complexity of preparing a  specific quantum state.
In general, quantifiers of complexity are scenario-dependent.
Here, we study  \emph{statistical} complexity of phase-space distributions for continuous-variable (CV) quantum states from an information-theoretic perspective.

In Ref.~\cite{Lopez-Ruiz1995}, it was proposed to  define a scalar value to capture the complexity of systems having a statistical description.
The main idea is to define complexity as the product of two scalar quantities, one that captures the spread of the distribution and one with the opposite behavior, i.e., which captures how peaked or ``far from equilibrium'' the distribution is.
The dynamical behavior of the complexity, introduced in Ref.~\cite{Lopez-Ruiz1995}, was studied in Ref.~\cite{Calb2001}.
This measure was also generalized to CV  distributions in Refs.~\cite{Catalan2002,Romera2004}.
Similar ideas involving only entropy have been introduced as well~\cite{Shiner1999}.
While these studies are limited to the classical case, a natural question arises: what about the quantum case?
And more specifically, what about the case of CV systems, where phase-space distributions are naturally defined?
In turn, characterizing and quantifying the inherent complexity of CV quantum states through phase-space distributions would offer deeper insights into their behavior and properties, along with the resources needed for their preparation.
 
The Wehrl entropy~\cite{Wehrl,Wehr1979}, derived from the Husimi function ($Q$-function)~\cite{Husi1940}, quantifies the uncertainty associated with the phase-space distributions of  quantum states. It is the quantum analogue of the differential entropy~\cite{Cove1991} and has been studied extensively for its important properties~\cite{Anis1976,Palm2018,Scut1979,Grab1984,Wehr1991,LiebW,LuoW,Miranowicz2001}.
It also offers insights into the quantum-classical correspondence and coherence properties of quantum states~\cite{Orlo1993,Lieb1994}, and plays a crucial role in quantifying many quantum features, such as squeezing~\cite{Lee1988}, entanglement~\cite{Mint2004,Marc2008}, non-Gaussianity~\cite{Ivan2012}, and nonclassicality~\cite{Zhan2021}.
The Wehrl entropy characterizes  how a quantum state is ``spread out'' in phase space, and can be interpreted as the ``classical'' information contained in CV quantum states, making it a potential tool for quantifying the complexity of such states.

The Fisher information, which is introduced to quantify ``the information amount of a probability distribution'' in statistical theory~\cite{Fish1925}, provides a significant way to measure the sensitivity of a statistical model to perturbations in its parameters~\cite{Cram1974,Stam1959,Frie1998}.
\FA{This concept can be extended to the quantum setting
leading to a whole family of quantum generalizations of the classical Fisher information~\cite{Yuen1973,Hels1976,Hole2011,Luo20031,Hornyak2007,Petz2010,Scandi2024a}.
These quantities have found several fundamental applications in analyzing various quantum phenomena and protocols~\cite{Luo2002,Brau1994,Luo20011,Luo2003,Sun2023,Zhon2013,Shitara2015a,Gess2018,Li2021}.
In this work, however, we focus on the phase-space properties of CV states and we consider the \emph{classical} Fisher information stemming from the Husimi $Q$-function.
Specifically, we will consider a quantity that operationally corresponds to the trace of the Fisher information matrix for displacement parameters.
This quantity can be understood as a functional of the CV state only, yet it is distinct from the family of quantum Fisher informations mentioned above.}

In this paper, we aim to explore the utility of \FA{the} Wehrl entropy and \FA{the} Fisher information \FA{(with respect to the location parameters of the Husimi function)} in quantifying the phase-space complexity of CV states. The remainder of the work is arranged as follows.
In Sec.~\ref{sec:complexity}, after reviewing some aspects of  the phase-space complexity, we introduce a quantifier of complexity in terms of Wehrl entropy and Fisher information based on the Husimi functions of quantum states. We reveal some basic properties of this quantifier.
In Sec.~\ref{sec:examples}, as illustrations, we evaluate the complexity of some typical quantum states including all Gaussian states and some non-Gaussian states.
In Sec.~\ref{sec:comparison}, we compare our quantifier of complexity of quantum states with a previous one, as well as some other quantifiers of important quantum features.
In Sec.~\ref{sec:generalization}, we introduce a family of alternative complexity quantifiers in terms of the $s$-ordered phase-space distributions.
Finally, we conclude with a summary and some discussions in Sec.~\ref{sec:summary}.
We will only treat the single-mode quantum states in this work.




\section{Complexity of CV states via Wehrl entropy and Fisher information}\label{sec:complexity}

To \mga{introduce} our study of complexity, we first consider complexity of classical systems described by probability distributions. For classical systems, \mga{a measure of statistical complexity of a 
probability vector $\boldsymbol{p}=(p_1,p_2,\cdots,p_n)$ can be defined as~\cite{Lopez-Ruiz1995}}
\begin{equation}
C_{\text{LMC}}(\boldsymbol{p}) = H (\boldsymbol{p}) D(\boldsymbol{p}),
\end{equation}
where $H(\boldsymbol{p})$ and $D(\boldsymbol{p})$  represent some measures of information and disequilibrium, respectively, of $\boldsymbol{p}$.
This definition captures the idea that complexity should vanish (or be minimal) both for perfect order and perfect disorder, e.g., both a perfect crystal at zero temperature and an ideal gas at equilibrium.

A possible way to generalize this idea to CV probability distributions $f(x)$ on $\mathbb{R}^d$ is the so-called Fisher-Shannon complexity quantifier~\cite{Romera2004}.
The main idea remains similar, however in this case the role of the measure of information is played by the Shannon entropy power
\begin{equation}
J(f) = \frac{1}{2 \pi e} e^{d S(f)/2},
\end{equation}
where $S(f) = - \int _{\mathbb{R}^d}f(x) \ln f(x)  \mathrm{d} x$ is the Shannon differential  entropy, which in general can also be negative, and $d$ is the dimension of the domain of $f$. On the other hand, the role of the  measure of disequilibrium is played by the Fisher information defined as
\begin{equation}
\label{eq:fisher_transl}
I(f) =\int _{\mathbb{R}^d} \frac{{\Vert {\nabla} f(x) \Vert^2}}{f(x)}\mathrm{d} x , 
\end{equation}
Where $\Vert {\nabla} f(x) \Vert^2 = {\nabla} f(x)^T {\nabla} f(x) = \sum_{j=1}^d [\partial_{x_j} f(x) ]^2 $ is the Euclidean norm of the gradient.
From an operational point of view, this quantity corresponds to the trace of the Fisher information matrix of the $d$ location parameters, i.e. the statistical model $f(x-\theta)$, where $\theta \in \mathbb{R}^d$.
For location parameters, the Fisher information does not depend on the value of $\theta$, which can be seen by a change of variables in the integral~\cite{Frie1998}.

The Fisher-Shannon measure of complexity is thus defined as~\cite{Romera2004}
\begin{equation}
C(f) = \frac{1}{d} J(f) I(f),
\end{equation}
where $I(f)$ and $J(f)$ are intimately connected, and represent respectively the ``surface area" and the ``volume" of the typical set associated to the probability distribution $f(x)$. The isoperimetric inequality for entropies \mga{implies} that $C(f) \geq 1,$ where the equality holds for multivariate normal probability distributions~\cite{Dembo1991}. Geometrically, there is an analogy between the fact that given a certain volume, balls (spheres) are the objects with minimal surface area and the fact that given a certain entropy, normal distributions are the ones with minimal Fisher information.

In the context of quantum systems, the generic distribution $f(x)$ has to be replaced by a quantum version. For a single quantum particle, it would just be the probability coming from the Born rule $f(x)=|\Psi(x)|^2$ for the wavefunction $\Psi (x).$ However, it has been observed that using the momentum or position density can yield different results, therefore different ways were proposed to treat every quadrature democratically~\cite{Manzano2012}.

On the other hand, since quantum states can be completely characterized via phase-space distributions, it is natural to consider those phase-space distributions as the quantum substitutes of $f(x)$. A similar idea has been explored~\cite{Hornyak2007} to define quantum Fisher information by using the class of phase-space probability distributions introduced by Cohen~\cite{Cohen1966}. However, when coming to the Shannon differential entropy, the phase-space distributions could cause troubles because they are not necessarily positive for all 
quantum states \mga{\cite{cerf,pizzimenti}.}
Hence, we choose a particular member from the phase-space distributions, the Husimi function, because it is nonnegative everywhere for every quantum state. Therefore, in this work, the complexity of a quantum state will be quantified by combining the Shannon differential entropy and the Fisher information defined in terms of the associated Husimi function.

Consider a single-mode CV quantum system described by the annihilation and creation operators $a$ and $a^\dagger$ satisfying the canonical commutation relation $$[a,a^\dagger ]={\bf 1}.$$ The displacement operators and the squeezing operators are defined as
\begin{align*}
D_{\xi}&=e^{\xi a^\dag -\xi^* a},\qquad \xi \in \mathbb{C},\\
S_{\eta} &= e^{\frac12 \eta (a^\dag)^2 -\frac12 \eta^* a^2},\qquad \eta \in \mathbb{C},
\end{align*}
respectively. They  play a fundamental role in \mga{CV quantum technology.}
The Husimi  function of a quantum state $\rho$ for such a system is defined as
\begin{equation}
\label{eq:Qdef}
Q(\alpha | \rho)=\langle \alpha | \rho | \alpha \rangle 
\end{equation}
and it is normalized as 
\begin{equation}
\label{eq:Qnorm}
\mga{\int_{\mathbb{C}} Q(\alpha | \rho)\!\frac{\mathrm{d}^{2}\alpha}{\pi}\, = 1}\,,
\end{equation}
where $\mathrm{d}^{2}\alpha=\mathrm{d}x \mathrm{d}y$ denotes the Lebesgue measure on $\mathbb{C}=\mathbb{R}^2, \alpha=x+iy, x,y \in \mathbb{R}$, and where $|\alpha\rangle$ are the coherent states, defined as the eigenstate of the annihilation operator: $a |\alpha\rangle = \alpha |\alpha\rangle$.
The Husimi function $Q(\alpha)$ contains the complete description of the quantum state (i.e., it is tomographically complete)~\cite{qsm,mul16} and may be sampled experimentally in a variety of systems~\cite{wc84,noh92,kir13}.

The Shannon differential entropy of $Q(\alpha | \rho)$ is the celebrated Wehrl entropy~\cite{Wehrl}
\begin{equation}
\label{eq:wehrl_entropy}
S_{\rm W} (\rho) = -\int_{\mathbb{C}}   Q(\alpha | \rho) \ln Q(\alpha | \rho) \frac{\mathrm{d}^{2}\alpha}{\pi}.
\end{equation}
The Wehrl entropy has many interesting properties. One of these is that it is minimal for single-mode CV states and equal to $1$ for coherent states. Despite being a type of Shannon differential entropy, the Wehrl entropy never becomes negative (actually possesses a positive lower bound due to the uncertainty \mga{relations} for the position and the momentum quadratures of the CV system), which is  due to the fact that the Husimi function cannot be too concentrated in phase-space in view of the uncertainty relations~\cite{LiebW,LuoW} and the corresponding added noise~\cite{yuen82,art88}.
Additionally, at fixed average energy, the Wehrl entropy is maximized by thermal states, like the von Neumann entropy~\cite{Orowski1999}.
In summary, the Wehrl entropy of the Husimi functions not only avoids negative values, but also captures important physical aspects of quantum states.


In the context of the Husimi function, the Fisher information defined by Eq.~\eqref{eq:fisher_transl} can be expressed as~\cite{Carlen1991,Luo2001}
\begin{equation}
\label{eq:fisher_husimi}
I (\rho) =\frac{1}{4} \int_{\mathbb{C}}  
\frac{{\Vert {\nabla} Q(\alpha | \rho) \Vert^2}}{Q(\alpha | \rho)} \frac{\mathrm{d}^{2}\alpha}{\pi},
\end{equation}
where the gradient operator are  with  respect to the real and imaginary parts of the complex number $\alpha$. Notice that we have introduced a factor $1/4$ for convenience 
compared to Eq.~\eqref{eq:fisher_transl}.
In this way, the Fisher information of the Husimi function of any \emph{pure} state is 1~\cite{Luo2001}, while for mixed 
states we have~\cite{Carlen1991,Luo2001},
\begin{equation}
I (\rho) \leq 1.
\end{equation}
Operationally, $I(\rho)$ is proportional to the trace of the Fisher information matrix for the problem of estimating the two parameters of a single-mode displacement~\cite{displa1,displa2,displa3,displa4}.

Let us now introduce a  quantifier of complexity of CV quantum states by employing both Wehrl entropy and Fisher information.

\vskip 0.2cm

 {\bfseries Definition 1.} The complexity of a quantum state $\rho$ is defined as
\begin{equation}
\label{eq:complexitypower_husimi}
{\cal C}(\rho)  =e^{S_{\rm W} (\rho )-1} I (\rho )
\end{equation}
with $S_{\rm W} (\rho )$ and $I(\rho )$ defined by Eqs. (\ref{eq:wehrl_entropy}) and (\ref{eq:fisher_husimi}), respectively.
\vskip 0.2cm
\mga{Throughout the paper, we will use the shorthand ${\cal C}(\psi)$  instead of ${\cal C}(\vert\psi\rangle\langle\psi\vert)$ for pure states $|\psi\rangle$. 
Notice that in  this case, $I (\psi) = 1$ and therefore ${\cal C}(\psi) = e^{S_{\rm W} (\psi )-1}$.}
Next, let us discuss some  important properties of this quantifier.

\vskip 0.2cm

{\bfseries Proposition 1.} The complexity ${\cal C}(\rho )$ is invariant under displacements and phase-space rotations in the sense that
\begin{align}
    {\cal C}(D_{\xi} \rho D_{\xi}^\dag) &= {\cal C}(\rho),\\
    {\cal C}(e^{i\phi a^\dag a}\rho e^{-i\phi a^\dag a})&={\cal C}(\rho),
\end{align}
where $D_{\xi}=e^{\xi a^\dag -\xi^* a},\ \xi \in \mathbb{C},$ are the displacement operators, and $e^{i\phi a^\dag a}, \phi \in [0,2\pi),$ are the rotation operators.

{\emph{Proof}}. Since the displacement operators and the phase-space rotations act as translations and rotations in the phase-space, respectively, i.e.,
$$Q (\alpha | D_{\xi} \rho D_{\xi}^\dag) = Q (\alpha-\xi | \rho),\qquad Q (\alpha | e^{i\phi a^\dag a}\rho e^{-i\phi a^\dag a}) = Q (e^{-i\phi}\alpha | \rho),$$ and the measure $\mathrm{d}^2\alpha / \pi$ is invariant under the displacements and rotations of the complex plane $\mathbb{C},$
they have no effect on the Fisher information or the Wehrl entropy of any state. Therefore, ${\cal C}(\rho )$ is invariant for these transformations. By contrast, the squeezing operators $S_{\eta} = e^{\frac12 \eta (a^\dag)^2 -\frac12 \eta^* a^2},\ \eta \in \mathbb{C}$ do not have a simple impact on the complexity, and their effects on complexity depend crucially on the initial states. For example, the complexity of the vacuum state will increase after it is squeezed, i.e., $${\cal C}(S_{\eta}|0\rangle)=\cosh|\eta| >{\cal C}(0)=1, \qquad \eta\neq0,$$
 but since $S_{\eta}^{-1}=S_{-\eta}$, it also holds that
 $${\cal C}(0)={\cal C}(S_{-\eta} S_{\eta}|0\rangle)<{\cal C}(S_{\eta}|0\rangle), \qquad \eta\neq0,$$
  that is, the complexity of the squeezed vacuum $S_{\eta}|0\rangle$ will decrease if it is further squeezed by $S_{-\eta}.$

\vskip 0.2cm

{\bfseries Proposition 2.} The complexity ${\cal C}(\rho)$ is invariant under a uniform phase-space scaling, in other words, if any two states $\rho_1$ and $\rho_2$ satisfy that
\begin{equation}\label{Cond}
Q (\alpha | \rho_1) = \lambda^2 Q(\lambda \alpha | \rho_2),
\end{equation}
where $\lambda>0$ is the scaling parameter, then we have
\begin{equation*}
{\cal C}(\rho_1)={\cal C}(\rho_2).
\end{equation*}

{\emph{Proof}}. For states $\rho_1$ and $\rho_2$ that satisfy the condition defined by Eq. (\ref{Cond}), direct calculation leads to
\begin{align}
    S_{\rm W} (\rho_1) &= -\ln (\lambda^2) + S_{\rm W} (\rho_2),\\
    I(\rho_1) &= \lambda^2 I(\rho_2).
\end{align}
Consequently, by Eq. (\ref{eq:complexitypower_husimi}), we have ${\cal C}(\rho_1)={\cal C}(\rho_2).$
\mga{This property will be useful in evaluating the complexity of photon-added thermal states, see Example 2 of 
Sec. 3.}

\vskip 0.2cm

{\bfseries Proposition 3.} The complexity ${\cal C}(\rho)$ \mga{satisfies} the following isoperimetric inequality
\begin{equation}
    \label{eq:isoperimetric}
    {\cal C}(\rho)\geq 1,\qquad  \forall \ \rho,
\end{equation}
and the equality holds if and only if $\rho$ is a (displaced) thermal state.

The proof follows directly from Ref.~\cite{Dembo1991}.
We remark that the vacuum $|0\rangle$ is a special case of thermal states (with zero thermal photon), and all coherent states achieve this lower bound as well.




\section{Examples}\label{sec:examples}

In this section, we evaluate the complexity of several classes of CV quantum states.
Our aim is twofold: to assess whether our measure aligns with the intuitive understanding of quantum state complexity, and to explore the relationships---and distinctions---between this notion of complexity and the various definitions of nonclassicality employed in CV  quantum technologies.


\subsection{Gaussian states}
\begin{figure}[ht!]
\centering
\includegraphics[width=.45\textwidth]{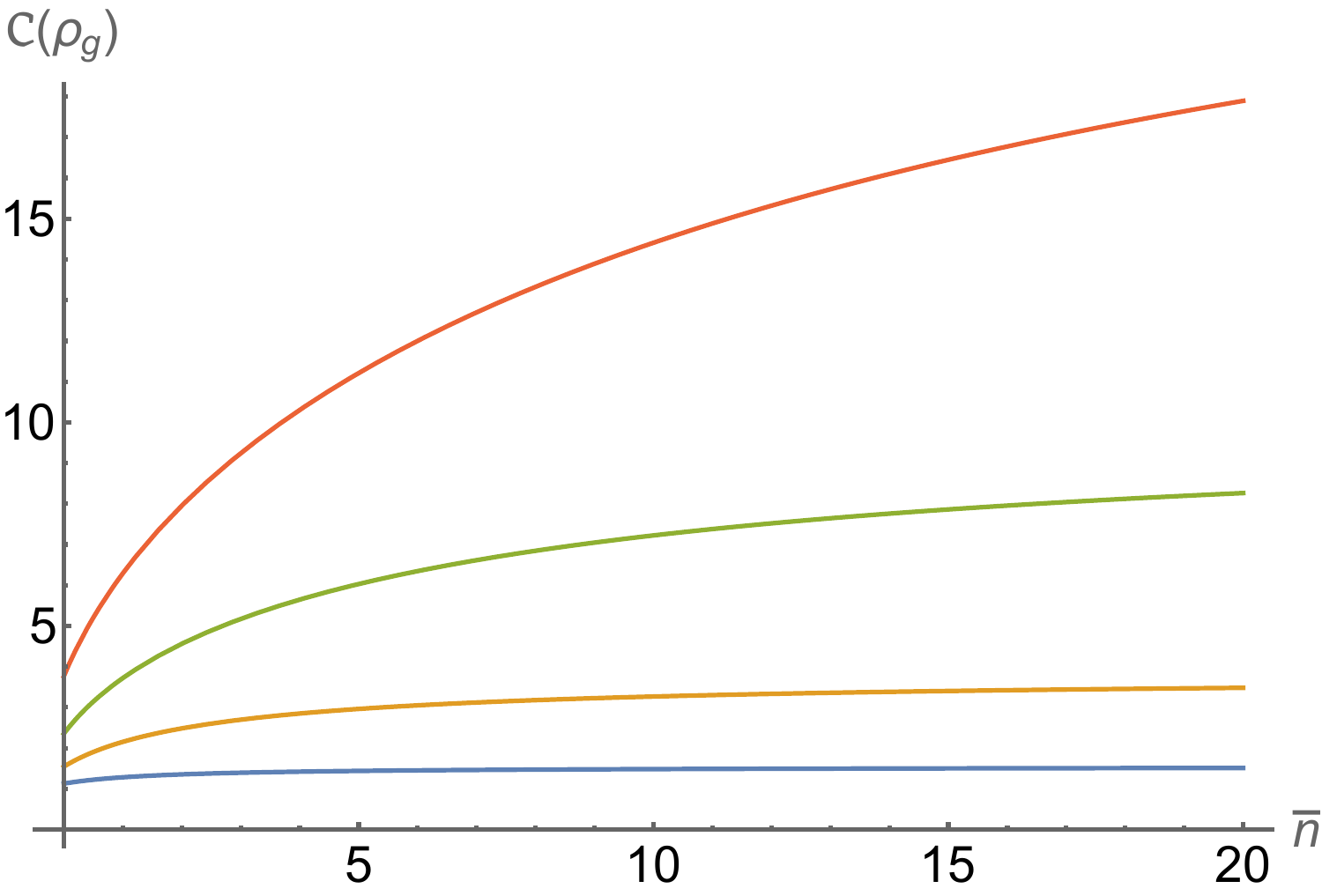}
\includegraphics[width=.45\textwidth]{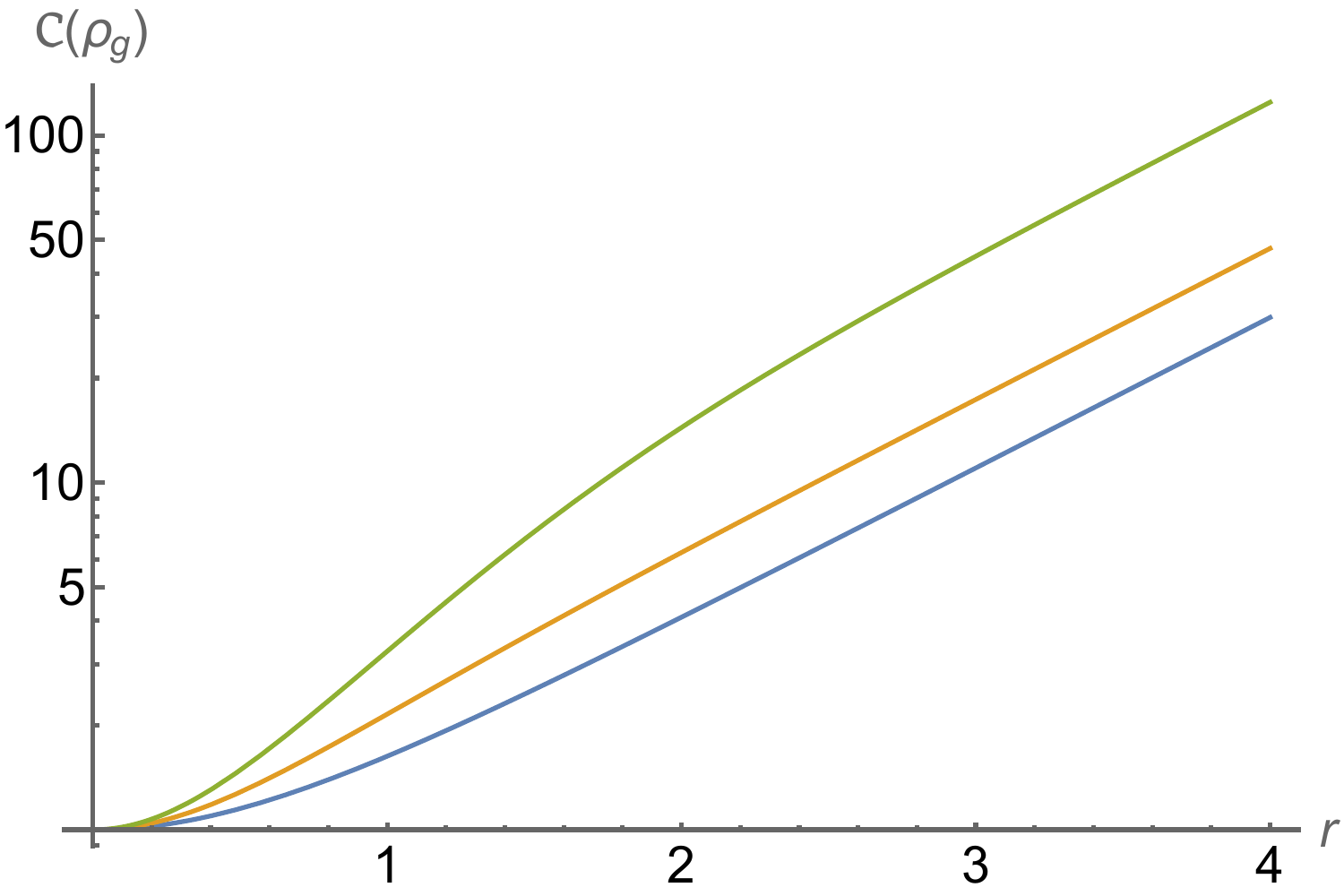}
\caption{(a) The complexity ${\cal C}( \rho_{\rm g} )$ of the Gaussian state $\rho_{\rm g}$ defined by Eq. (\ref{eq:gaussian_complexity_power}) as a function of $\bar{n}$ with $r=0.5,1,1.5,2$ (from bottom to top). (b) Semilog plot of complexity ${\cal C}( \rho_{\rm g} )$ as a function of $r$ with $\bar{n}=0.1,1,10$ (from bottom to top).}
\label{fig1}
\end{figure}

 Any single-mode Gaussian state can be expressed as a displaced squeezed thermal state in the form
\begin{equation}
    \label{eq:def_gaussian_state}
    \rho_{\rm g} =D_{\xi} S_{\eta} \rho_{\mathrm{th}} S_{\eta}^\dag D_{\xi}^\dag, \qquad \xi,  \eta \in \mathbb{C},
\end{equation}
where
\begin{equation}\label{thermal}
\rho_{\mathrm{th}}=\frac{1}{\bar{n}+1}\sum_{n=0}^\infty \left(\frac{\bar{n}}{\bar{n}+1}\right)^n |n\rangle\langle n|, \qquad \bar n\geq 0
\end{equation}
is a thermal state, and $\bar n = \mathrm{tr}(\rho_{\mathrm{th}} a^\dag a)\geq 0$ is the average thermal photon number. The Husimi function of this  Gaussian state reads
\begin{equation}
Q(\alpha | \rho_{\rm g} )= \frac{1}{\sqrt{\Delta}} e^{\cyan{\left( -A|\alpha-\xi|^2 + B \left( 
e^{-i\theta} (\alpha-\xi)^2 + e^{i\theta} (\alpha^*-\xi^*)^2 \right) \right)}/2\Delta},\qquad \alpha\in\mathbb C,
\end{equation}
where $\eta=r e^{i\theta},\ r\geq 0, \ \theta\in [0,2\pi),$ and
\begin{align}
    \Delta &= (\bar{n}+1)^2 +(2\bar{n}+1)\sinh^2 r,\label{Delta} \\
   A &= \cyan{1+(2\bar{n}+1)\cosh(2r),} \label{A}\\
   B &=  \cyan{\left (\bar{n}+\frac12\right )\sinh(2r)} \label{B}.
\end{align}
By direct calculation, the Wehrl entropy and the Fisher information can be evaluated as
\begin{align}
\label{eq:gaussian_wehrl}
S_{\rm W} ( \rho_{\rm g} ) &= 1+ \frac{1}{2} \ln\Delta, \\	
I ( \rho_{\rm g} ) &=\frac{\cyan{A}}{2\Delta}.\label{eq:gaussian_fisher}
\end{align}
We see that they do not depend on the displacement parameter $\xi$ and the squeezing angle $\theta.$ 
Evidently, the Fisher information is exactly 1 when $\bar{n}=0$ and also tends to 1 as we fix $\bar{n}$ and increase the squeezing parameter $r$ to infinity. This observation highlights that even if the Fisher information is constant for pure states, it is not solely a function of the purity of the Gaussian states (which only depends on $\bar{n}$).
Now, we readily obtain the complexity of the Gaussian state $\rho_{\rm g} $ as
\begin{equation}
\label{eq:gaussian_complexity_power}
{\cal C}(\rho_{\rm g} ) = \frac{\cyan{A}}{2\sqrt\Delta},
\end{equation}
where $\Delta$ \cyan{and $A$} are defined by Eqs.~(\ref{Delta}) \cyan{and (\ref{A}), respectively}.
To gain an intuitive understanding of this quantity, we plot the curves of the complexity of Gaussian state ${\cal C}(\rho _{\rm g})$ with respect to the parameter $\bar{n}$ for fixed $r=0.5,1,1.5,2,$ on the left panel of Fig.~\ref{fig1}, and with respect to the parameter $r$ for fixed $\bar{n}=0.1,1,10,$ on the right panel of Fig.~\ref{fig1}. 
The complexity ${\cal C}( \rho_{\rm g} )$ tends to $\cosh (2r)$ as $\bar{n}$ tends to infinity with $r$ fixed, and tends to $+\infty$ as $r$ goes to infinity.

Consider an energy constraint given by
\begin{equation}
\label{eq:meanphotons}
E=\mathrm{tr}(\rho_{\rm g}  a^\dag a) = \bar{n}+|\xi|^2+(2\bar{n}+1)\sinh^2 r.
\end{equation}
Under this constraint, we aim to maximize the complexity.

\vskip 0.2cm

{\bfseries Proposition 4.} Among all Gaussian states $\rho_{\rm g} $  defined by Eq. (\ref{eq:def_gaussian_state}) with a fixed average photon number $E=\mathrm{tr} (\rho_{\rm g}  a^\dag a)$, the most complex states are the squeezed vacuum states, while the least complex states are the displaced thermal states.

{\emph{Proof}}. First, note that ${\cal C}( \rho_{\rm g} )$ is an increasing function of $r,$ in order to maximize ${\cal C}( \rho_{\rm g} ),$ we should set $\xi=0,$ which is obvious from Eq. (\ref{eq:meanphotons}). The constraint then becomes
\begin{equation*}
\cosh(2r)=\frac{2E+1}{2\bar{n}+1}.
\end{equation*}
Substituting this into the expression of ${\cal C}( \rho_{\rm g} )$, i.e. Eq. (\ref{eq:gaussian_complexity_power}), we obtain
\begin{equation*}
    {\cal C}( \rho_{\rm g} )=\frac{E+1}{\sqrt{\bar{n}(\bar{n}+1)+E+1}},
\end{equation*}
which is a decreasing function of $\bar{n}.$ Thus ${\cal C}( \rho_{\rm g} )$ is maximized at $\bar{n}=0$, i.e., the maximal complexity ${\cal C}( \rho_{\rm g} ) =\sqrt{E+1}$ is achieved by the squeezed vacuum state $S_{\eta}|0\rangle$, with $r= |\eta| = \ln\big(\sqrt{E}+\sqrt{E+1}\big)$.

On the other hand, the minimal complexity is of course ${\cal C}( \rho_{\rm g} )=1$ when $\eta=0$, which corresponds to all  displaced thermal states with $\bar{n}+|\xi|^2=E$.


\subsection{Photon-added thermal states}

A $k$-photon-added thermal state
$\rho_{k}\propto (a^\dag)^k\rho_{\mathrm{th}} a^k,k=1,2,\cdots,$ can be expressed as
\begin{equation}
    \rho_{k}=\frac{1}{k! (\bar{n}+1)^{k+1}} \sum_{n=0}^\infty \left(\frac{\bar{n}}{\bar{n}+1}\right)^n (n+k)\cdots(n+1)|n+k\rangle\langle n+k|,
\end{equation}
where $\rho_{\mathrm{th}}$ is the thermal state defined by Eq. (\ref{thermal}), and $\bar n = \mathrm{tr}(\rho_{\mathrm{th}} a^\dag a)\geq 0$ is the average thermal photon number. Its Husimi function is
\begin{equation}
    Q (\alpha | \rho_{k})     = \frac{|\alpha|^{2k}}{k!(\bar{n}+1)^{k+1}} e^{-\frac{|\alpha|^2}{\bar{n}+1}} ,\qquad \alpha\in\mathbb C.
\end{equation}
Note that the Husimi  function of the Fock state $|k\rangle$ is
\begin{equation}
    Q (\alpha | k) = \langle\alpha|k\rangle\langle k|\alpha\rangle = \frac{|\alpha|^{2k}}{k!} e^{-|\alpha|^2}, \qquad \alpha\in\mathbb C.
\end{equation}
Hence $Q (\alpha | \rho_{k})$ is a scaled version of $Q (\alpha | k),$ that is,
\begin{equation}
    Q (\alpha | \rho_{k})=\frac{1}{\bar{n}+1} Q \Big (\frac{\alpha}{\sqrt{\bar{n}+1}} \Big| k \Big ).
\end{equation}
By Proposition 2, we obtain that $ {\cal C}(\rho_{k})= {\cal C}(k)$, which implies that the complexity of the $k$-photon-added thermal states are independent of the temperature, similar to the case of thermal states.
To calculate the complexity explicitly, we note that since the Fock state $|k\rangle$ is pure and
\begin{equation}
    \label{eq:Fock_complexity}
    {\cal C}(\rho_{k})= {\cal C}(k) = e^{S_{\rm W} (k)-1} = k!e^{k-k\psi(k+1)},
\end{equation}
where the last equality follows from the Wehrl entropy of the Fock state~\cite{Orowski1999}:
\begin{equation}
    S_{\rm W} (k) = 1+k+\ln(k!)-k\psi(k+1).
\end{equation}
Here $\psi(k+1)=\sum_{m=1}^k \frac{1}{m}-\gamma$ is the digamma function and $\gamma \simeq 0.577$ is the Euler constant.
As expected, the complexity increases as more photons are added into a thermal state.


\subsection{Photon-added coherent states}

A photon-added coherent state is defined as
\begin{equation}
\label{eq:phot_add_coh}
|\psi_{\text{pac}} \rangle = \frac{1}{\sqrt{1 + |\beta|^2}} a^{\dag} |\beta\rangle, \qquad  \beta \in \mathbb{C}
\end{equation}
and the corresponding Husimi function is~\cite{Agarwal1991}
\begin{equation}
\label{eq:Qpacs}
Q(\alpha | \psi_{\text{pac}})= \frac{|\alpha|^2}{1+|\beta|^2} e^{- |\alpha-\beta|^2}, \qquad \alpha\in\mathbb C.
\end{equation}
\begin{figure}[h!]
    \centering
    \includegraphics[width=0.5\linewidth]{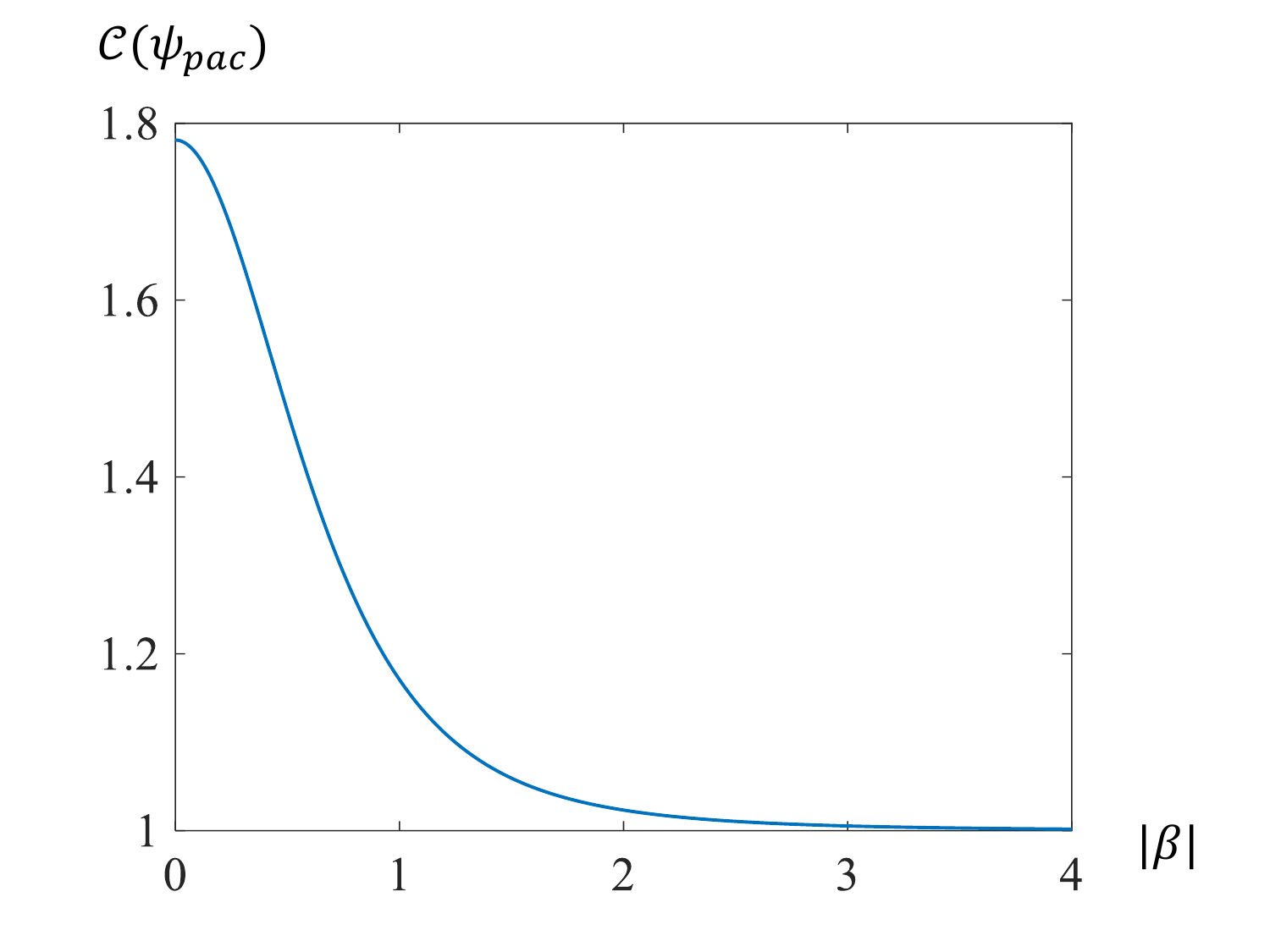}
    \caption{The complexity ${\cal C}(\psi _{\rm pac} )$ of the photon-added coherent state $|\psi _{\rm pac}\rangle $ as a function of the amplitude $\beta$. \mga{The complexity does not depend on the phase of $\beta$. 
    }}
    \label{fig3}
\end{figure}

The complexity ${\cal C}(\psi_{\text{pac}})$ \mga{does not depend on the argument of $\beta$ and can be 
evaluated} numerically. Results are shown in Fig.~\ref{fig3}, \mga{where  ${\cal C}(\psi_{\text{pac}})$ is depicted as a function of $|\beta|$. As it is apparent from the plot, complexity  decreases with  $|\beta|$,} which can be \mga{understood by the following argument: a photon-added coherent state can be rewritten as}
\begin{align}
    \nonumber
    |\psi_{\text{pac}} \rangle &=\frac{1}{\sqrt{1 + |\beta|^2}} a^{\dag} D_{\beta}|0\rangle
   = \frac{1}{\sqrt{1 + |\beta|^2}} D_{\beta} (a^{\dag}+\beta^*)|0\rangle\\
    &= D_{\beta} \left(\frac{\beta^*}{\sqrt{1+|\beta|^2}} |0\rangle + \frac{1}{\sqrt{1+|\beta|^2}} |1\rangle \right),
\end{align}
and, as proved in Proposition 1, displacements do \cyan{not} change the complexity. \mga{Therefore, the complexity of 
$|\psi_{\text{pac}} \rangle$ is equal to the complexity of a superposition of the vacuum $|0\rangle$ and the Fock state $|1\rangle$, with the superposing coefficients determined by the amplitude $\beta$. We thus have that}
${\cal C}(\psi_{\text{pac}}) \to {\cal C}(1)=e^\gamma$ as $|\beta| \to 0$, where $\gamma$ is the Euler constant, while ${\cal C}(\psi_{\text{pac}} ) \to {\cal C}(0)=1$ as $|\beta| \to \infty$.


\subsection{Cat states}

Consider the cat state
\begin{equation}
|\psi_{\text{cat}} \rangle = \frac{1}{N_{\beta}}\big ( |\beta\rangle + e^{i \phi } |-\beta	 \rangle \big ), \qquad \beta \in \mathbb{C}, \phi\in[0,2\pi), \label{CAT}
\end{equation}
with the normalization constant $N_{\beta}=\sqrt{2\left(1+ e^{-2|\beta|^2}\cos\phi\right)}$. For $\phi = 0,$ we have an even-cat state, for $\phi = \pi,$ we have an odd-cat state, and for $\phi = \frac{\pi}{2},$ we have a Yurke-Stoler cat state.
The Husimi function of this state reads

\begin{equation}
\label{eq:Qcat_phi}
Q(\alpha | \psi_{\text{cat}})=\frac{1}{N^2_{\beta}} \Big( e^{-|\alpha-\beta|^2} + e^{-|\alpha+\beta|^2}  + 2 e^{- |\alpha|^2 - |\beta |^2}\cos\big(\phi + \cyan{2} {\rm Im} (\alpha \beta^*) \big) \Big),\qquad \alpha\in\mathbb C.
\end{equation}
Consistently with Ref.~\cite{Buek1995}, we find that states with $\phi=\pi$ have higher Wehrl entropy, and thus higher complexity, while the states with $\phi=0$ have the minimal entropy.
Actually this is not surprising, because the relative phase is also connected to the energy
\begin{equation}
\label{eq:energycats}
E\left(\phi \right) ={\rm tr}(|\psi_{\text{cat}} \rangle\langle\psi_{\text{cat}}|a^\dag a)= |\beta|^2 \frac{1- e^{-2|\beta|^2}\cos\phi}{1+ e^{-2|\beta|^2}\cos\phi }.
\end{equation}
Therefore a higher value of complexity and Wehrl entropy corresponds to a higher energy.

\begin{figure}[h!]
\centering
\includegraphics[width=.7\textwidth]{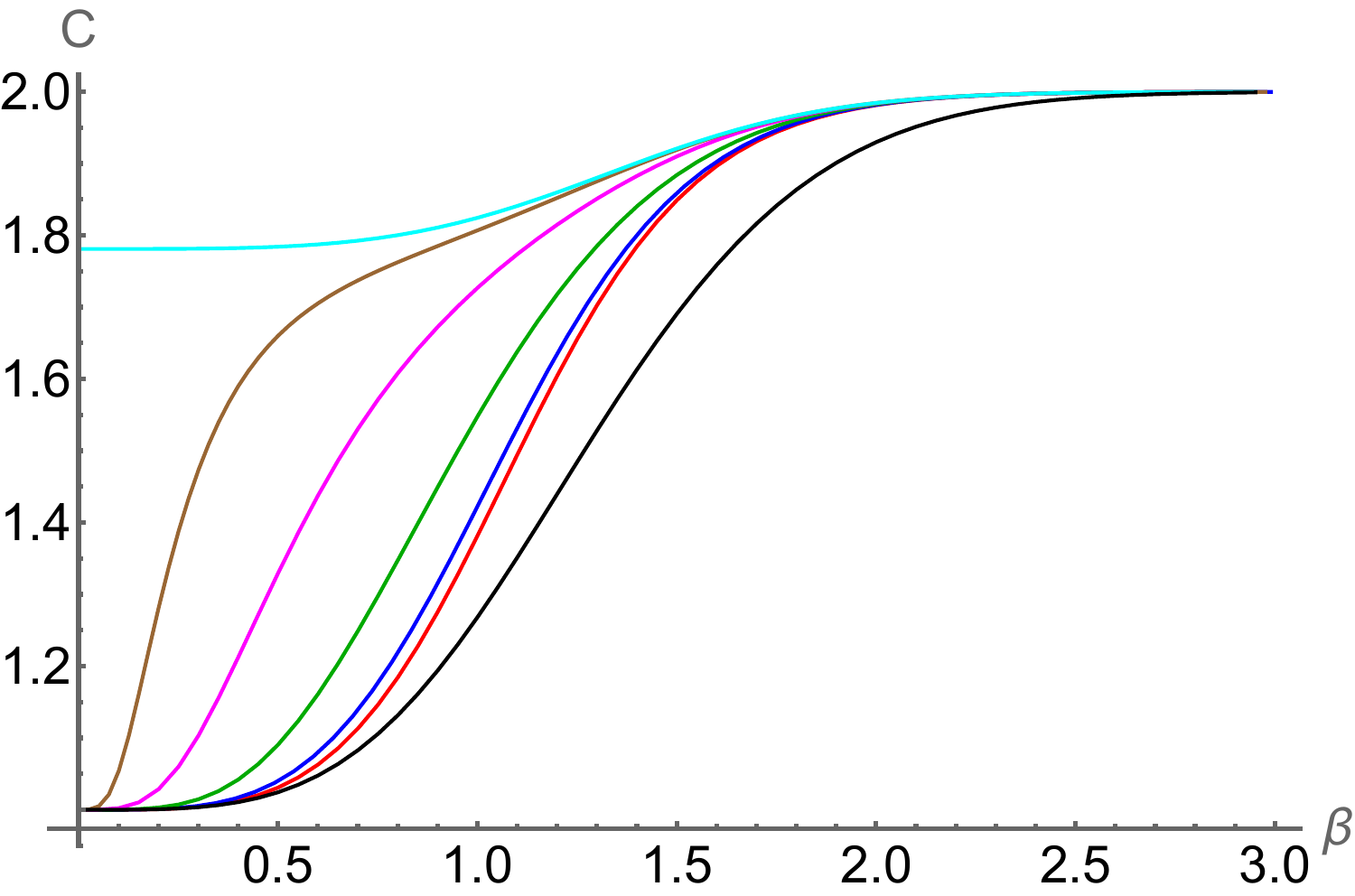}
\caption{\cyan{The complexity ${\cal C}(m_\beta)$ of a mixture of two coherent states (black line) versus the complexity ${\cal C}(\psi_{\rm cat})$ of cat states (colored lines) as functions of the coherent state amplitude $\beta$. We show results for different values of the relative phase $\phi$ defined in Eq. (\ref{CAT}): $\phi=0, \frac{\pi}{4}, \frac{\pi}{2}, \frac{3\pi}{4},\frac{9\pi}{10},\pi$ (from bottom to top). The odd-cat state is not well defined as $\beta \to 0.$}}
\label{fig4}
\end{figure}

In this context, it is interesting to compare the complexity of superpositions versus mixtures of coherent states. The problem of distinguishing a superposition of coherent states from a mixture by means of the Wehrl entropy was analyzed in Ref.~\cite{Buek1995}. It is noted that when the component coherent states are ``far away'', the Husimi  function of the superposition becomes approximately equal to the Husimi  function of the corresponding mixture, independently of the relative phase. It has been pointed out that other distributions derived from the Husimi function might be more suitable to distinguish coherent superpositions from mixtures~\cite{Miranowicz2001}.
In our framework, we investigate whether the inclusion of the Fisher information term in the complexity quantifier we defined allows for a  better distinction between superpositions and mixtures. Intuitively, the complexity of a mixture should be smaller, as the Fisher information of the Husimi  function is less than one for mixed states.
Consider the following mixture
\begin{equation}
\label{eq:cat_mixed}
m_{\beta} = \frac{1}{2} |\beta \rangle \langle \beta| + \frac{1}{2} |-\beta\rangle\langle - \beta |, \qquad \beta \in \mathbb{C}
\end{equation}
and its Husimi  function is simply the sum of the Husimi functions corresponding to the two coherent states.
In Fig.~\ref{fig4}, we compare the complexity ${\cal C}(m_\beta)$ of the mixture with  that of the cat state $|\psi _{\rm cat}\rangle $  defined by Eq.~\eqref{CAT}, as a function of the amplitude $\beta$, \cyan{and we assume here $\beta \in \mathbb{R}^+$}. The complexity ${\cal C}(m_\beta)$ of the mixture $m_\beta$ is always lower, but beyond  a certain amplitude $|\beta |$, the difference becomes negligible. This behavior is not surprising, considering the results discussed in Ref.~\cite{Buek1995}. 

\section{Comparison with other quantifiers}\label{sec:comparison}
We first compare our complexity ${\cal C}(\rho)$ with the complexity quantifier proposed by Manzano~\cite{Manzano2012}, which is defined as a product of the Fisher information and the entropic power of Shannon entropy. Recall that instead of the phase-space function $Q(\alpha|\rho)$, in Manzano's approach to complexity, the Fisher information and the entropy are defined in terms of the one-dimensional probability distribution obtained by measuring the observable $\cos (\omega) X - \sin (\omega) P$, where $X$ and $P$ are the position and momentum observables, respectively. Therefore, this complexity quantifier depends on the basis of measurement, which is captured by the parameter $\omega \in [0,\pi)$. Two ways are  proposed to eliminate the dependence on a specific basis: one involves averaging over $\omega$, while the other involves taking the minimum. In both cases, the complexity quantifier reaches its minimum value for all Gaussian states.

The most significant difference between our quantifier of complexity and Manzano's lies in that, while the latter is  based on one-dimensional probability distributions, ours is based on the phase-space distributions, which are two-dimensional in nature. The difference can be illustrated by Gaussian states. If there is no squeezing, then a Gaussian state has a bivariate normal distribution with equal variances in the position and momentum quadratures. Such states will have the minimum complexity ${\cal C}(\rho)=1$, and they are exactly the displaced thermal states. However, when there is a nontrivial squeezing, the variances will be unequal and hence do not meet the equality condition of the two-dimensional isoperimetric inequality. Therefore their complexity will be strictly greater than 1. However, when restricted to
one-dimension, regardless of the chosen basis, Gaussian states will always have a normal distribution, thus satisfying the equality condition of the one-dimensional isoperimetric inequality. Intuitively, squeezing should increase complexity, which suggests that it is important to consider a joint phase-space distribution in this context. We summarize the above comparisons in Table \ref{tab:comp_complexity}.
\begin{table}[!h]
    \centering
    \begin{tabular}{|c|c|c|} \hline
         Quantifier of complexity &  (Quasi)probability distribution
&  Minimum achieved by\\ \hline
         Manzano's  ${\cal M}(\rho )$ &  marginal
&  Gaussian states\\ \hline
         Ours ${\cal C}(\rho )$ &  joint &  (displaced) thermal states\\ \hline
    \end{tabular}
    \caption{Comparison between two different quantifiers of complexity: ${\cal M}(\rho )$ vs. ${\cal C}(\rho )$.}
    \label{tab:comp_complexity}
\end{table}

Next, we analyze the relationship of ${\cal C}(\rho)$ with some quantifiers for nonclassicality and non-Gaussianity in the literature.

Mandel's $\mathcal{Q}$ factor~\cite{Mandel1979}
\begin{equation}
    \mathcal{Q}(\rho) = \frac{\text{tr}(\rho a^{\dag 2} a^2) - \big(\text{tr}(\rho a^{\dag} a)\big)^2}{\text{tr}(\rho a^{\dag} a)}
\end{equation}
is the first quantifier for nonclassicality, which characterizes the deviations from  the Poissonian photon statistics of states. More specifically, $\mathcal{Q}(\rho)>0$, $\mathcal{Q}(\rho)=0$, and $\mathcal{Q}(\rho)<0$ correspond to photon distributions being super-Poissonian, Poissonian, and sub-Poissonian, respectively. From the definition, we can easily show that Mandel's $\mathcal{Q}$ factor is lower bounded by $-1$, and this lower bound can be achieved by all Fock states: $\mathcal{Q} (k)=-1, \ \forall \ k$. Mandel's $\mathcal{Q}$ factor for Gaussian states defined by Eq. (\ref{eq:def_gaussian_state}) can be calculated straightforwardly as
\begin{equation}
    \label{eq:mandelQ_gaussian}
    \mathcal{Q} (\rho_{\rm g} ) = \frac{ (\bar{n}+\frac12)^2\cosh (4r) + 2(\bar{n}+\frac12)|\xi \cosh r + \xi^* e^{i\theta} \sinh r|^2  - \frac14}{ (\bar{n}+\frac12)\cosh (2r) + |\xi|^2 - \frac12} - 1.
\end{equation}
While our complexity quantifier depends only on the thermal photon number $\bar{n}$ and the squeezing parameter $r$ of the Gaussian states, this quantifier depends additionally on the displacement parameter $\xi$ and the squeezing angle $\theta$.

The nonclassical depth is defined as~\cite{Lee1991,Lee1992}
\begin{equation}
    \label{eq:def_nc_depth}
    \tau_m(\rho) = \frac12 \Big ( 1-\sup_{-1\leq s \leq 1} \{s: W_s(z|\rho) \geq 0 ,\ \forall z\in\mathbb{C}\} \Big ),
\end{equation}
where $W_s(z|\rho)$ is the $s$-ordered phase-space distribution introduced by Cahill and Glauber in a seminal work of phase-space representation of quantum states~\cite{Cahill1969}. For a fixed \cyan{$s\leq 1$}, 
\begin{equation}
    \label{eq:s-ordered-quasi}
    W_{s} (\alpha | \rho) = \int_{\mathbb{C}}   e^{\alpha \xi^* - \alpha^* \xi} \chi_s (\xi | \rho) \frac{\mathrm{d}^{2} \xi}{\pi}, \qquad \alpha \in \mathbb{C},
\end{equation}
which is the symplectic Fourier transform of the $s$-ordered characteristic function
\begin{equation}
    \chi_s (\xi | \rho) = e^{s|\xi|^2/2} \text{tr} (\rho D_{\xi} ),\qquad \xi \in \mathbb{C}.
\end{equation}
The special cases where $s=-1,0,1$ correspond to the Husimi function, Wigner function, and Glauber-Sudarshan $P$-function, respectively.
The nonclassical depth $\tau_m(\rho)$ describes the minimum amount of thermal noise to make the singular $P$-function nonnegative. By definition, we have $0\leq \tau_m(\rho) \leq 1$, and a greater value of $\tau_m(\rho)$ indicates a more nonclassical state.
By calculation, the nonclassical depth of the Gaussian states defined by Eq. (\ref{eq:def_gaussian_state}) is
\begin{equation}
    \label{eq:nonclassical_depth_gaussian}
    \tau_m(\rho_{\rm g} )=\max \Big\{0, \frac{(\bar{n}+1)\tanh{r}-\bar{n}}{1+\tanh{r}} \Big \}.
\end{equation}
The original definition requires $\tau_m(\rho) \geq 0$. However, a negative value of $\tau_m(\rho)$ would also make some sense, because it can be interpreted as the degree of classicality. Therefore we may modify the definition and remove the floor to define 
\begin{equation}
    \label{eq:gaussian_nc_depth}
    \tilde{\tau}_m(\rho_{\rm g} )=\frac{(\bar{n}+1)\tanh{r}-\bar{n}}{1+\tanh{r}}.
\end{equation}
For instance,  $\tilde{\tau}_m (\rho_\text{th}) = -\bar{n}$, which may be interpreted as indicating that the thermal states are more classical than coherent states.

From Eqs.~\eqref{eq:gaussian_complexity_power} and~\eqref{eq:gaussian_nc_depth}, we see that the two quantifiers, ${\cal C}(\rho_{\rm g})$ and $\tilde {\tau}_m (\rho_{\rm g}),$ 
capture different aspects of Gaussian states. The complexity ${\cal C}(\rho_{\rm g} )$ increases in the thermal photon number $\bar{n},$ while the nonclassical depth $\tilde{\tau}_m(\rho_{\rm g} )$ decreases in $\bar{n}$. Another important difference lies in their bounds. For the complexity,  ${\cal C}(\rho_{\rm g} )\geq1$, giving a lower bound on the complexity of Gaussian states, and from the expression we can see that for the other direction, the Gaussian states can be as complex as needed. In contrast, for the nonclassical depth, $\tilde{\tau}_m(\rho_{\rm g} ) < \frac12$, giving an upper bound on the maximal degree of nonclassicality.
Moreover, sometimes the nonclassicality depth looks like an all-or-nothing indicator, and for many quite different states, it turns out to be the same value. For example, all Fock states achieve the maximal nonclassical depth ${\tau}_m(k)=1, \forall \ k$, but our complexity quantifier shows that those with larger photon numbers are more complex. \mga{In turn, this is capturing the idea that more complex physical interactions are required to generate Fock states with higher $k$.}

Another quantifier of nonclassicality is defined in terms of the Wigner-Yanase skew information as~\cite{Luo2019}
\begin{equation}
    {\cal N}(\rho) = -\frac12 \text{tr} ([\sqrt{\rho},a]^2).
\end{equation}
For the Gaussian states $\rho_{\rm g}$ defined by Eq. (\ref{eq:def_gaussian_state}), we have
\begin{equation}
   {\cal  N}(\rho_{\rm g} ) = \Big ( \frac12+\bar{n}-\sqrt{\bar{n}(\bar{n}+1)} \Big )\cosh(2r).
\end{equation}
Note that for fixed $r$, it decreases in $\bar{n}$ as well. For Fock states, we have
\begin{equation}
  {\cal  N}(k) =\frac12+k.
\end{equation}

Wigner negativity is also a well-known quantifier for nonclassicality, which is defined as the integrated negative part of the Wigner function~\cite{Kenfack2004}, i.e.,
\begin{equation}
    \delta(\rho) = \int_{\mathbb{C}}   |W_0 (z| \rho)| \frac{\mathrm{d}^{2}z}{\pi} -1.
\end{equation}
\FA{We mention that this notion of nonclassicality also has a resource-theoretical interpretation~\cite{Albarelli2018,Takagi2018}.}  
For Gaussian states, their Wigner functions are positive definite, which lead to
\begin{equation}
    \delta(\rho_{\rm g} )=0.
\end{equation}
For other quantum states the Wigner negativity is usually not easy to calculate. For Fock states, analytical results are only available for the first few ones~\cite{Kenfack2004}, and for other Fock states, there are only numerical results.

Finally, let us compare ${\cal C}(\rho)$ with two quantifiers of non-Gaussianity. One is the Hilbert-Schmidt distance from a Gaussian reference~\cite{Genoni2007,Genoni2010}
\begin{equation}
    \delta_A (\rho) = \frac{ \text{tr} (\rho-\sigma)^2}{2\text{tr}(\rho^2)},
\end{equation}
and the other is the quantum relative entropy to a Gaussian reference state~\cite{Genoni2008,Genoni2010}
\begin{equation}
    \delta_B (\rho) = \text{tr} \big( \rho(\ln \rho - \ln \sigma) \big),
\end{equation}
where in both definitions, $\sigma$ is the  Gaussian reference state with the same mean vector and covariance matrix as $\rho$.

Since both quantifiers characterize the degree of non-Gaussianity, naturally, they vanish for Gaussian states
\begin{equation}
    \delta_A (\rho_{\rm g} ) = \delta_B (\rho_{\rm g} ) = 0.
\end{equation}
Therefore, it is immediately clear that our quantifier of complexity is distinct from a quantifier of non-Gaussianity in nature

The above comparison is summarized in Table \ref{tab:comp_quantifier}.
\begin{table}
    \centering
    \begin{tabular}{|l|l|l|} \hline
         Quantifier  &  Gaussian state $\rho_{\rm g} $& Fock state $|k\rangle$\\ \hline
         Complexity ${\cal C}(\rho)$&  Eq. (\ref{eq:gaussian_complexity_power})& $k!e^{k-k\psi(k+1)}$\\ \hline
         Mandel's factor $\mathcal{Q}(\rho)$& Eq. (\ref{eq:mandelQ_gaussian}) & $-1$ \\\hline
         Nonclassical depth $\tau_m(\rho)$& $\max \{0, \frac{(\bar{n}+1)\tanh{r}-\bar{n}}{1+\tanh{r}} \}$ & 1\\ \hline
         Wigner-Yanase skew information ${\cal N}(\rho)$& $\big ( \frac12+\bar{n}-\sqrt{\bar{n}(\bar{n}+1)} \big )\cosh(2r)$ & $\frac12+k$\\ \hline
         Wigner negativity $\delta(\rho)$&  0& numerical results\\ \hline
         Hilbert-Schmidt distance $\delta_A(\rho)$ &  0& $\frac12 \big ( 1+\frac{1}{2k+1} \big ) - \frac{k^k}{(k+1)^{k+1}}$ \\ \hline
         Quantum relative entropy $\delta_B(\rho)$&  0& $(k+1)\ln(k+1) - k\ln k$ \\ \hline
    \end{tabular}
    \caption{Comparison among quantifiers for different quantum features.}
    \label{tab:comp_quantifier}
\end{table}




\section{Generalization to the $s$-ordered complexity}\label{sec:generalization}

In this section, we generalize our complexity quantifier ${\cal C}(\rho )$ to  other phase-space distributions. At first, we notice that while the Husimi function is nonnegative everywhere, i.e., $Q(\alpha | \rho) \geq 0$ for any quantum state $\rho$, in general the $s$-ordered quasiprobability distribution $W_{s} (\alpha | \rho)$ could be negative somewhere for some states, and a state-dependent threshold on $s$ should be introduced.  

\vskip 0.2cm

 {\bfseries Definition 2.} The $s$-ordered complexity of an optical state $\rho$ is defined in terms of its $s$-ordered phase-space distribution as
\begin{equation}
\label{eq:complexitypower_sorder}
{\cal C}_s (\rho) =e^{S_{{\rm W}_{\!s}} (\rho) - 1} I_s (\rho),
\end{equation}
where
\begin{align}
    S_{{\rm W}_{\!s}} (\rho) &= -\int_{\mathbb{C}}     W_{s} (\alpha | \rho) \ln W_{s} (\alpha | \rho) \frac{\mathrm{d}^{2} \alpha}{\pi},\\
    I_s (\rho) &= \frac{1}{4} \int_{\mathbb{C}}    \frac{||{\nabla} W_{s} (\alpha | \rho)||^2}{W_{s} (\alpha | \rho)}  \frac{\mathrm{d}^{2}\alpha}{\pi}.
\end{align}
In particular, when $s=-1$, ${\cal C}_{-1} (\rho)$ reduces to the complexity ${\cal C}(\rho)$ introduced by Definition 1.
When $W_{s} (\alpha | \rho)$ is nonnegative everywhere, $S_{{\rm W}_{\!s}} (\rho)$ and $I_s (\rho)$ are the Shannon differential entropy and the Fisher information of $W_{s} (\alpha | \rho)$. In order to elucidate some points here we explore three classes of classical states: coherent states $|\beta\rangle$, thermal states $\rho_{\mathrm{th}}$ defined by Eq. (\ref{thermal}), and phase-averaged coherent states
\begin{equation}\label{rhob}
\rho_{\beta} = \int_{0}^{2\pi}\!\! 
\frac{\mathrm{d} \theta}{\pi}  |\beta\rangle \langle \beta |\,, \qquad \beta =|\beta|e^{i\theta}
\end{equation}
obtained by averaging the phase of coherent states $|\beta \rangle.$ 
We remark that these states are also diagonal on the number state basis since
\begin{equation}
    \rho_{\beta} = e^{-|\beta|^2} \sum_{n=0}^\infty \frac{|\beta|^{2n}}{n!} |n\rangle\langle n|.
\end{equation}
These three classes of states are all classical, and their $s$-ordered phase-space distributions can be directly evaluated as
\begin{align}
\label{eq:qp_coherent}
W_{s} (\alpha | \beta)&=\frac{2}{1 -s} e^{ - \frac{2}{1 - s} |\alpha- \beta|^2},\qquad \alpha\in\mathbb{C},\\
\label{eq:qp_thermal}
W_{s} (\alpha | \rho_{\text{th}})&=\frac{2}{2 \bar{n} +1 -s}e^{ - \frac{2}{2 \bar{n} +1 - s} |\alpha|^2},\qquad\alpha \in\mathbb{C},\\
\label{eq:qp_phase_av_coh}
W_{s} (\alpha | \rho_{\beta})&=\frac{2}{1 -s} e^{ - \frac{2}{1 - s} \left( |\alpha|^2+ |\beta|^2 \right)} I_0 \left(\frac{4 |\alpha\beta|}{1-s} \right),\qquad \alpha\in\mathbb{C},
\end{align}
where $I_0(\cdot)$ is the modified zeroth-order Bessel function of the first kind.

\begin{figure}[h!]
\centering
\includegraphics[width=.5\textwidth]{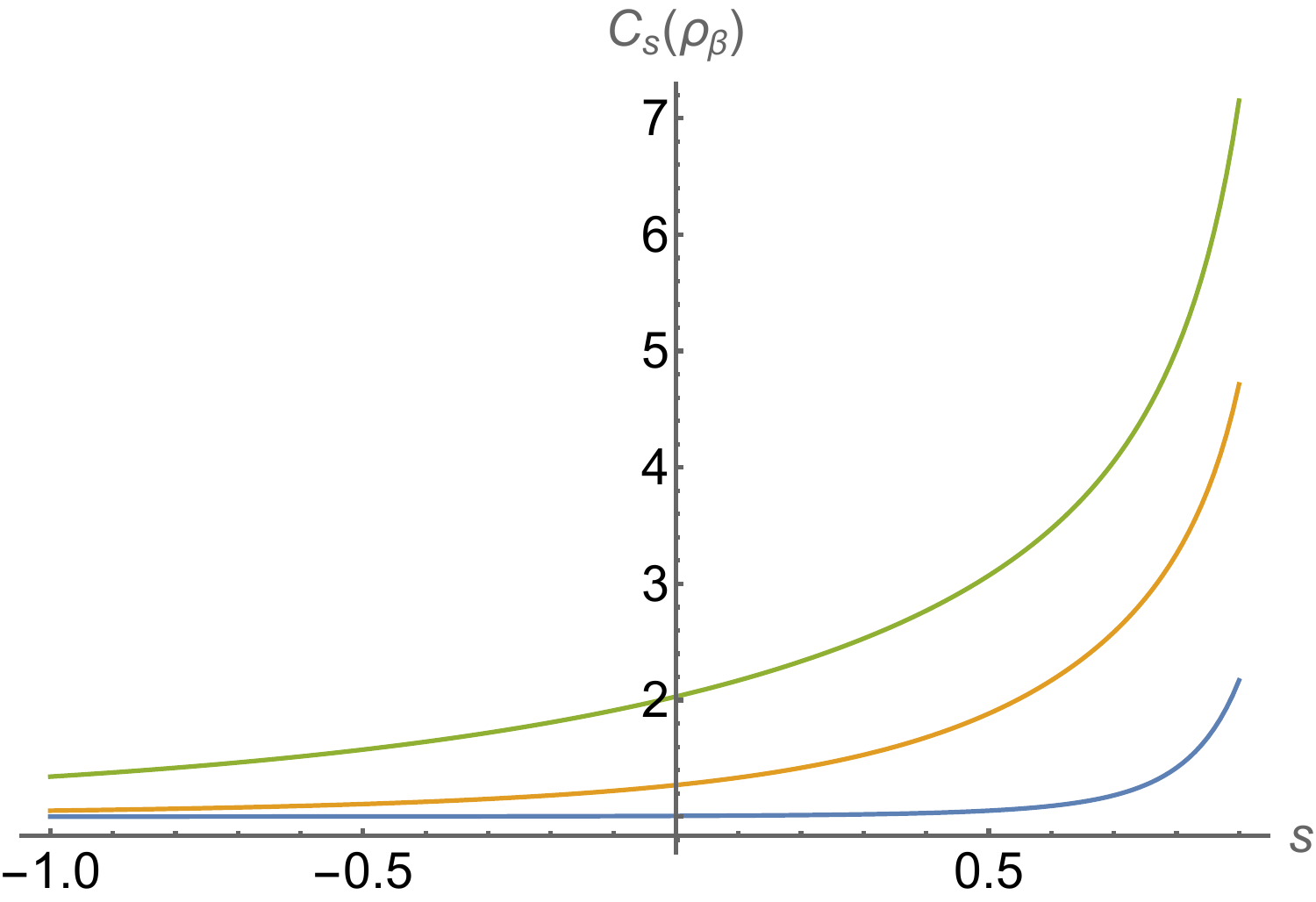}
\caption{\cyan{The $s$-ordered complexity ${\cal C}_s(\rho_{\beta})$ of the phase-averaged coherent state $\rho_{\beta}$ as a function of the ordering $s$, with $|\beta|=0.5,1,1.5$ (from bottom to top).}}
\label{fig:phase_averaged}
\end{figure}

For the coherent states characterized by the distributions $W_{s} (\alpha | \beta),$ as $s \to 1$ the differential entropy $S_{{\rm W}_{\!s}} (\beta) \to -\infty$, so the entropy power $e^{S_{{\rm W}_{\!s}} (\beta) -1} \to 0$, while the Fisher information $I_s (\beta) \to +\infty$. However, ${\cal C}_s (\beta) = e^{S_{{\rm W}_{\!s}} (\beta) -1} I_s (\beta) = 1$, which is independent of $s.$ Therefore for  any value of the ordering $s$, the statistical complexity of coherent states remains to be the minimal value 1.

For the thermal states, we have exactly the same result and the complexity remains to be $1$ for any ordering. No limiting procedure is needed, as the 
$P$-functions are Gaussian. The results for coherent states and thermal states are actually just a consequence of the isoperimetric inequality, as their distributions are bivariate normals for any ordering.

For the phase-averaged coherent states $\rho_{\beta},$ in general the $s$-ordered phase-space distributions defined by Eq.~\eqref{eq:qp_phase_av_coh} are non-Gaussian~\cite{Allevi2013}. 
\cyan{In Fig.~\ref{fig:phase_averaged} we plot the $s$-ordered complexity $\mathcal{C}_s (\rho_{\beta} )$ of the phase-averaged coherent states with respect to the ordering parameter $s$ for fixed $\beta=0.5,1,1.5$. We see that the a larger value of $s$ gives a more sensitive measure of complexity in the sense that }the complexity quantifier is an increasing function of the physical amplitude $|\beta|$, but by going from the Husimi  distribution towards the $P$-function, the complexity rises more quickly.

\cyan{Another interesting observation is that if we rescale the amplitude $|\Tilde{\beta}|=|\beta|/\sqrt{1-s}$, then} the complexity quantifier \emph{does not} depend on the ordering $s$. The proof of the invariance property is easy. After rescaling, the $s$-ordered quasiprobability distributions become
\begin{equation}
    W_{s}(\alpha | \rho_{\beta})=\frac{2}{1 -s} e^{-2 |\Tilde{\beta}|^2 - \frac{2 |\alpha|^2}{1 - s} } I_0 \left(\frac{4 |\alpha| |\Tilde{\beta}|}{\sqrt{1-s}} \right),\qquad \alpha\in\mathbb C.
\end{equation}
Then different orderings $s$ just correspond to different scalings. So the same proof of Proposition 2 applies. This is true as long as we consider the other parameters of the state fixed, however the rescaling is also applied to other parameters of the distribution. In the case of the coherent states, this is not important, as the results do not depend on the displacement. However in the case of the phase average coherent states, the change of ordering also affects the amplitude parameter $\beta$, so in order to observe the same behaviour of the complexity, we need to use the rescaled amplitude by ${|\beta|}/{\sqrt{1-s}}$.
This property of the complexity under rescaling of variables of the phase-space distribution has been already observed~\cite{Romera2004}. \mga{In future study, it would be interesting to investigate
the precise conditions for which 
changing the ordering of the phase-space distribution amounts only to a rescaling of the argument 
of the distribution. In mathematical terms, this means knowing for which functions Gaussian convolution only amounts to a rescaling of variables.}

\cyan{We can also derive the $s$-ordered complexity for some nonclassical states. For any single-mode Gaussian state $\rho_{\rm g}$ defined by Eq. (\ref{eq:def_gaussian_state}), whenever the ordering satisfies $s<1 - 2 \tau_m (\rho_{\rm g})$, where $\tau_m (\rho_{\rm g})$ is the nonclassical depth given by Eq. (\ref{eq:nonclassical_depth_gaussian}), the $s$-ordered quasiprobability distribution is a well-defined Gaussian function~\cite{Zhang2022}
\begin{equation}
    \label{eq:quasi_gaussian}
    W_s (\alpha | \rho_{\rm g}) = \frac{1}{\sqrt{\Delta_s}} e^{\left( -A_s|\alpha-\xi|^2 + B \left( 
e^{-i\theta} (\alpha-\xi)^2 + e^{i\theta} (\alpha^*-\xi^*)^2 \right) \right)/2\Delta_s},\qquad \alpha\in\mathbb C,
\end{equation}
where
\begin{align}
    \Delta_s &= \left(\bar{n}+ \frac{1-s}{2} \right)^2 -s(2\bar{n}+1)\sinh^2 r,\\
   A_s &= -s+(2\bar{n}+1)\cosh(2r), 
\end{align}
and $B$ is defined as in Eq. (\ref{B}) which does not depend on $s$. 
Following our definition, we have
\begin{eqnarray}
    S_{{\rm W}_{\!s}} (\rho_{\rm g}) &=& 1 + \frac12 \ln \Delta_s, \\
    I_s (\rho_{\rm g}) &=& \frac{A_s}{2\Delta_s}.
\end{eqnarray}
Then the $s$-ordered complexity of the Gaussian state $\rho_{\rm g}$ reads
\begin{equation}
    \mathcal{C}_s (\rho_{\rm g} ) = \frac{A_s}{2\sqrt{\Delta_s}}.
\end{equation}
}

\begin{figure}[h!]
    \centering
    \includegraphics[width=0.55\linewidth]{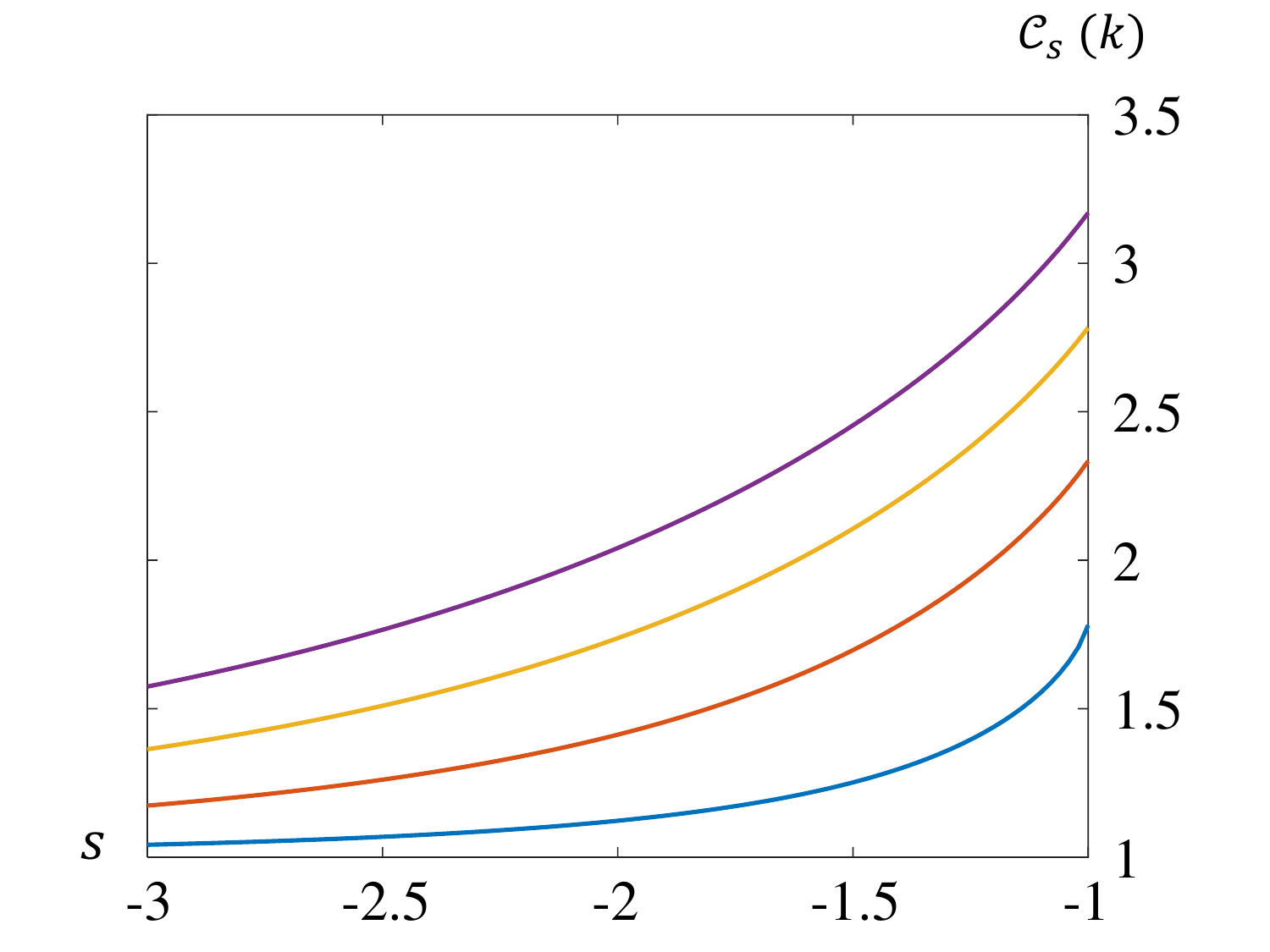}
    \caption{The $s$-ordered complexity ${\cal C}_s(k)$ of the Fock state $|k\rangle$ as a function of the ordering $s$, with $k=1,2,3,4$ (from bottom to top).}
    \label{fig:f5_Fock_s}
\end{figure}

Another interesting example is the Fock state $|k\rangle$. Of course, for $s>-1$ the quasiprobability function $W_s (\alpha | k)$ will have negative values. So we focus on $s<-1$. In this case, $W_s (\alpha | k)$ is a convolution of the Husimi function $Q(\alpha | k )$ with Gaussian noise~\cite{Cahill1969}, i.e.,
\begin{equation}
    W_s (\alpha | k) = \frac{2}{-1-s} \int_\mathbb{C} Q(\beta | k ) e^{-\frac{2|\alpha-\beta|^2}{-1-s}} \frac{\mathrm{d}^{2}\beta}{\pi},
\end{equation}
which is nonnegative everywhere, $\forall \ s<-1$. The explicit form of the above function has been calculated in Ref.~\cite{Lee1991}, as
\begin{equation}
    W_s (\alpha | k) = \frac{2}{1-s} \left( \frac{-1-s}{1-s} \right)^k e^{-\frac{2}{1-s} |\alpha|^2} L_k \left( \frac{4|\alpha|^2}{1-s^2} \right),
\end{equation}
where $L_k(\cdot)$ is the Laguerre polynomial.
Then we can evaluate numerically the $s$-ordered complexity ${\cal C}_s (k)$. In Fig.~\ref{fig:f5_Fock_s} we plot ${\cal C}_s (k)$ as a function of $s$ for the first few $k$. We see that ${\cal C}_s(k)$ tends to 1 as $s$ goes to minus infinity, which makes sense because we are adding too much thermal noise so that the complexity approaches the complexity of a thermal state, which is always 1.

\cyan{In this example, we also observe that a smaller value of $s$ corresponds to a less sensitive measure of complexity, because $\mathcal{C}_s (k) \to 1$ as $s \to -\infty$, regardless of $k$, which means that the different Fock states will become indistinguishable. This is consistent with the observation we made for the phase-averaged coherent state example. Therefore, in order to have a more sensitive measure we want to choose a larger value of $s$. However when $s>-1$ we may encounter negative values in the quasiprobability distributions. Overall, this justifies our choice of $s=-1$, i.e., the Husimi function, as the optimal one to introduce a sensible measure of complexity.}


\section{Summary and discussion}\label{sec:summary}

Complexity is a \mga{rather} subtle aspect of quantum states, and we cannot expect a single quantity to capture all \FA{its} features.
It is desirable to study it from different angles.
\FA{By using the} Wehrl entropy (which characterizes  delocalization) and the Fisher information (which characterizes metrological power and \FA{phase space localization}) of the Husimi functions of  quantum states, we have introduced a quantifier of complexity of \FA{CV} quantum states.
We have revealed some basic properties of the quantifier.
We have analytically evaluated the quantifier for various states including general Gaussian states, photon-added thermal states, classical states, etc.
We have illustrated the behaviors of the quantifier under variation of various parameters.
Moreover, we have compared the quantifier with some other quantifiers of important quantum features.
Finally, we have introduced a family of alternative complexity quantifiers in terms of the $s$-ordered phase-space distributions.

Since the operational meaning and significance  of both the Wehrl entropy and Fisher information are well established, we expect our complexity quantifier to find applications in quantum information tasks, e.g., in operationally quantifying physical resources required to generate quantum states with a given complexity.
Moreover, since  complexity intermingles both classical and quantum features, it may play an interesting role in studying the classical-quantum boundary, which is a key issue in quantum information theory.

\FA{In this work,} we have focused on single-mode CV quantum states.
It would be natural to extend our definition to multi-mode states. For example, consider the case of two-mode states $\rho_{12}$ with Husimi function $Q(\alpha_1,\alpha_2|\rho_{12})=\langle\alpha_1,\alpha_2|\rho_{12}|\alpha_1,\alpha_2\rangle ,$ where $|\alpha_1,\alpha_2\rangle=|\alpha_1\rangle\otimes|\alpha_2\rangle, \alpha_1,\alpha_2\in\mathbb C,$ are the two-mode coherent states. Based on this, the phase-space complexity of two-mode states can be defined as ${\cal C}(\rho_{12})  = e^{ \frac12 S_{\rm W} (\rho_{12} )-1} I (\rho_{12} )$, where $S_{\rm W} (\rho_{12} )$ and $I (\rho_{12} )$ are defined similarly as in Eqs. (\ref{eq:wehrl_entropy}) and (\ref{eq:fisher_husimi}), with the Husimi function $Q(\alpha | \rho)$ replaced by $Q(\alpha_1,\alpha_2|\rho_{12})$ and the measure $\frac{\mathrm{d}^{2}\alpha}{\pi}$ replaced by $\frac{\mathrm{d}^{2}\alpha_1}{\pi} \frac{\mathrm{d}^{2}\alpha_2}{\pi}$. Then again by the isoperimetric inequality~\cite{Dembo1991}, we have ${\cal C}(\rho_{12}) \geq 2.$ From the definition, the quantifier ${\cal C}(\cdot)$ is not additive for product states: $S_{\rm W} (\cdot)$ and $I(\cdot)$ are indeed additive in the sense that $S_{\rm W} (\rho_{12})=S_{\rm W} (\rho_1)+S_{\rm W} (\rho_2)$ and $I(\rho_{12})=I(\rho_1)+I(\rho_2)$ if $\rho_{12} = \rho_1 \otimes
\rho_2$.
However, ${\cal C}(\cdot)$ is not additive as there are cross terms because of the multiplication.
Thus correlations come into play as an essential ingredient of complexity, a feature that correctly captures our intuition. This multi-partite extension and its correlating aspects present interesting challenges and will be investigated elsewhere.
\section*{Acknowledgements}
MGAP thanks the Chinese Academy of Sciences (CAS) for supporting a visit to AMSS in Beijing, where part of this work was conducted.  This work was supported by  the National Key R\&D Program of China, Grant No. 2020YFA0712700, and the National Natural Science Foundation of China, Grant Nos.  12426671 and 12341103.

\nocite{apsrev4Control}
\bibliography{cvcomplx-with-doi,revtex-custom}

\end{document}